\input{aipcheck}

\documentclass[,final,sort&compress]{aipproc}

\layoutstyle{8x11single}
   \usepackage{graphicx}
   \usepackage{epstopdf}
  \usepackage{amssymb}

\def\teq#1{$\, #1\,$}                         
\def\apj{ApJ}

\def\app{Astroparticle Phys.}                   
\def\apss{Astr. Space Sci.}                     
\def\asr{Adv. Space Res.}                       

\def\mnras{{M.N.R.A.S.}}
\def\prl{Phys. Rev. Lett.}                      
\def\prd{Phys. Rev. D}                          
\def\ssr{Space Sci. Rev.}                       
\newcommand{\vol}[2]{$\;$\bf #1\rm , #2.}           
             \font\sevenrm=cmr7

\def\erg{\varepsilon_\gamma}
\def\thetascatt{\theta_{\hbox{\sevenrm scatt}}}
\def\thetaBone{\Theta_{\hbox{\sevenrm Bf1}}}

\def\betaHTone{\beta_{\hbox{\sevenrm 1HT}}}

\begin{document}
\begin{flushright}
\phantom{p}
\vspace{-60pt}
     To appear in Proc. of the {\it 6th Huntsville Gamma-Ray Burst Symposium} (2009),\\ 
     eds. C.~A. Meegan, N. Gehrels, \& C. Kouveliotou (AIP Conf. Proc., New York).
\end{flushright} 

\newcommand{\figureoutpdf}[5]{\centerline{}
   \centerline{\hspace{#3in} \includegraphics[width=#2truein]{#1}}
   \vspace{#4truein} \caption{#5} \centerline{} }

\title{Probes of Diffusive Shock Acceleration using\\ Gamma-Ray Burst Prompt Emission}

\author{Matthew G. Baring}{
    address={Department of Physics and Astronomy, MS-108,
                      Rice University, P. O. Box 1892, \\
                      Houston, TX 77251-1892, USA\hskip 10pt
                      {\rm Email: baring@rice.edu}}
}

\keywords{Gamma-ray bursts, non-thermal emission; 
diffusive shock acceleration; hydromagnetic turbulence}

\classification{98.70.Rz; 95.85.Pw; 98.70.Sa; 52.35.Ra; 
52.25.Xz; 52.27.Ny; 52.35.Tc; 52.65.Pp}

\begin{abstract}
The principal paradigm for gamma-ray bursts (GRBs) suggests that the
prompt transient gamma-ray signal arises from multiple shocks internal
to the relativistic expansion.  This paper explores how GRB prompt
emission spectra can constrain electron (or ion) acceleration properties
at the relativistic shocks that pertain to GRB models. The array of
possible high-energy power-law indices in accelerated populations is
highlighted, focusing on how spectra above 1 MeV can probe the field
obliquity in GRB internal shocks, and the character of hydromagnetic
turbulence in their environs.  When encompassing the MeV-band spectral
break, fits to BATSE/EGRET burst data indicate that the preponderance of
electrons responsible for the prompt emission reside in an intrinsically
non-thermal population.  This differs markedly from typical populations
generated in acceleration simulations; potential resolutions of this
conflict such as the action of self-absorption are mentioned.  Spectral
modeling also suggests that the synchrotron mechanism is favored over
synchrotron self-Compton scenarios due to the latter's typically broad
curvature near the peak. Such diagnostics will be enhanced by the
broadband spectral coverage of bursts by the {\it Fermi} Gamma-Ray Space
Telescope; the GBM will provide key information on the lower energy
portions of the non-thermal particle population, while the LAT will
constrain the power-law regime of particle acceleration.  
\end{abstract}

\maketitle


\section{Introduction}
 \label{sec:Introduction}
The gamma-ray burst paradigm has evolved dramatically over the last
decade since the first redshift determinations provided unambiguous
evidence for a cosmological distance scale.  Yet much is still to be
learned about associations, progenitors, and the environment of both the
prompt and afterglow emission regions, and the physical processes active
therein. For prompt burst emission, the measurement of the high energy
spectral index provides a key constraint on the interpretation of the
electron acceleration process.  Presuming the popular burst paradigm of
radiative dissipation at internal shocks that accelerate particles
\cite{RM92,Piran99,Meszaros02}, it is of great interest to understand
what physical conditions in the shocked environs can elicit the observed
indices and the spectral structure around the MeV-band peak.  To assess
this, one turns to models of diffusive shock acceleration in
relativistic systems.  These come in different varieties, including
semi-analytic solutions of the diffusion-convection equation
\cite{KH89,Kirk00}, Monte Carlo simulations
\cite{EJR90,ED04,NO04,SBS07}, and particle-in-cell (PIC) full plasma
simulations \cite{Hoshino92,Nishikawa05,Medvedev05,Spitkovsky08}. Each
has its merits and limitations.  Tractability of the analytic approaches
generally restricts solution to power-law regimes for the particle
momentum distributions.  PIC codes are rich in their information on
turbulence and shock-layer electrodynamics. To interface with GRB data,
a broad dynamic range in momenta is desirable, and this is the natural
niche of Monte Carlo simulation techniques, the focus of this paper.

A key property of acceleration at the relativistic shocks that are
presumed to seed prompt GRB radiative dissipation is that the
distribution functions \teq{f(\hbox{\bf p})} are inherently anisotropic.
This renders the power-law indices and other distribution
characteristics sensitive to directional influences such as the magnetic
field orientation, and the nature of MHD turbulence that often
propagates along the field lines. Accordingly, familiar results from
relativistic shock acceleration theory such as the so-called canonical
\teq{\sigma=2.23} power-law index \cite{Kirk00}, while providing useful
insights, are of fairly limited applicability.  This is fortunate since
the GRB database so far has exhibited a substantial range of spectral
characteristics, so any universal signature in the acceleration
predictions would be unnecessarily limiting. This paper explores some of
the features of diffusive shock acceleration using results from a Monte
Carlo simulation, and addresses probes of the theoretical parameter
space imposed by extant GRB observations around and above the MeV
spectral break.  Palpable diagnostics \cite{BB04} have already been
enabled by data provided by the BATSE and EGRET instruments on the
Compton Gamma-Ray Observatory (CGRO) for a handful of bright bursts. The
burst community expects significant advances in the understanding of
such constraints in the next few years, afforded by the broad spectral
coverage and sensitivity of the GBM and LAT experiments on NASA's {\it
Fermi} Gamma-Ray Space Telescope.

\section{Diffusive Acceleration at Relativistic Shocks}
 \label{sec:DSA}
Most internal GRB shocks that are putatively responsible for the prompt
emission are mildly-relativistic because they are formed by the
collision of two ultra-relativistic shells.  Hence, this regime forms
the focus of this exposition.  The discussion will address the
theoretical methodology employed, and the resulting characteristics of
the diffusive acceleration process, before moving onto constraints on
theoretical parameters imposed by the prompt GRB observations.

The simulation used here to calculate diffusive acceleration in
relativistic planar shocks is a kinematic Monte Carlo technique that has
been employed extensively in supernova remnant and heliospheric
contexts, and is described in detail in numerous papers
\cite{EJR90,JE91,EBJ95,ED04,SB06,BS09}. It is conceptually similar to
Bell's \cite{Bell78} test particle approach to diffusive shock
acceleration.  Particles are injected upstream and gyrate in a laminar
electromagnetic field, with their trajectories being governed by a
relativistic Lorentz force equation in the frame of the shock.  In
general, the fluid frame magnetic field is inclined at an angle
\teq{\thetaBone} to the shock normal. Because the shock is moving with a
velocity {\bf u}({\bf x}) relative to the plasma rest frame, there is,
in general, a {\bf u $\times$ B} electric field in addition to the bulk
magnetic field.  Particle interactions with Alfv\'{e}n wave and other
hydromagnetic turbulence is modeled by using a phenomenological
scattering of the charges in the rest frame of the plasma.  The
scattering precipitates spatial diffusion of particles along magnetic
field lines, and to a varying extent, across them as well.  The
scatterings are also assumed to be quasi-elastic, an idealization that
is usually valid because in most astrophysical systems the flow speed
far exceeds the Alfv\'{e}n speed, and contributions from stochastic
second-order Fermi acceleration are small. The diffusion permits a
minority of particles to transit the shock plane numerous times, gaining
energy with each crossing via the shock drift and first-order Fermi
processes.

A continuum of scattering angles, between large-angle (LAS) or
small-angle cases, can be modeled by the simulation.  In the local fluid
frame, the time, \teq{\delta t_f}, between scatterings is coupled
\cite{EJR90} to the mean free path, \teq{\lambda}, and the maximum
scattering (i.e. momentum deflection) angle, \teq{\thetascatt} via \teq{
\delta t_f\approx \lambda\thetascatt^{2}/(6v)} for particles of speed
\teq{v\approx c}.  Usually \teq{\lambda} is assumed to be proportional
to a power of the particle momentum \teq{p} (see \cite{EJR90,GBS92} for
microphysical justifications for this choice), and for simplicity it is
presumed to scale as the particle gyroradius, \teq{r_g}, i.e.
\teq{\lambda=\eta r_g\propto p}. The parameter \teq{\eta} in the model
is a measure of the level of turbulence present in the system, coupling
directly to the amount of cross-field diffusion, such that \teq{\eta =1}
corresponds to the isotropic {\it Bohm diffusion} limit, where the field
fluctuations satisfy \teq{\delta B/B\sim 1}.  In kinetic theory,
\teq{\eta} couples the parallel (\teq{\kappa_{\parallel}=\lambda v/3})
and perpendicular (\teq{\kappa_{\perp}}) spatial diffusion coefficients
via the relation \teq{\kappa_{\perp}/\kappa_{\parallel}=1/(1+\eta^{2})}
\cite{FJO74,EBJ95}.  In parallel shocks, where the {\bf B} field is
directed along the shock normal (\teq{\thetaBone=0}), \teq{\eta} has
only limited impact on the resulting energy spectrum, principally
determining the diffusive spatial scale normal to the shock.  However,
in oblique relativistic shocks where \teq{\thetaBone > 0}, the diffusive
transport of particles across the field (and hence across the shock)
becomes critical to retention of them in the acceleration process. 
Accordingly, for such systems, the interplay between the field angle and
the value of \teq{\eta} controls the spectral index of the particle
distribution \cite{ED04,Baring04}, a feature that is central to the
interpretation of GRB spectra below.

\subsection{Acceleration Properties at Relativistic Shocks}
 \label{sec:accel_prop}
\begin{figure}
 \centerline{
  \includegraphics[width=.51\textwidth]{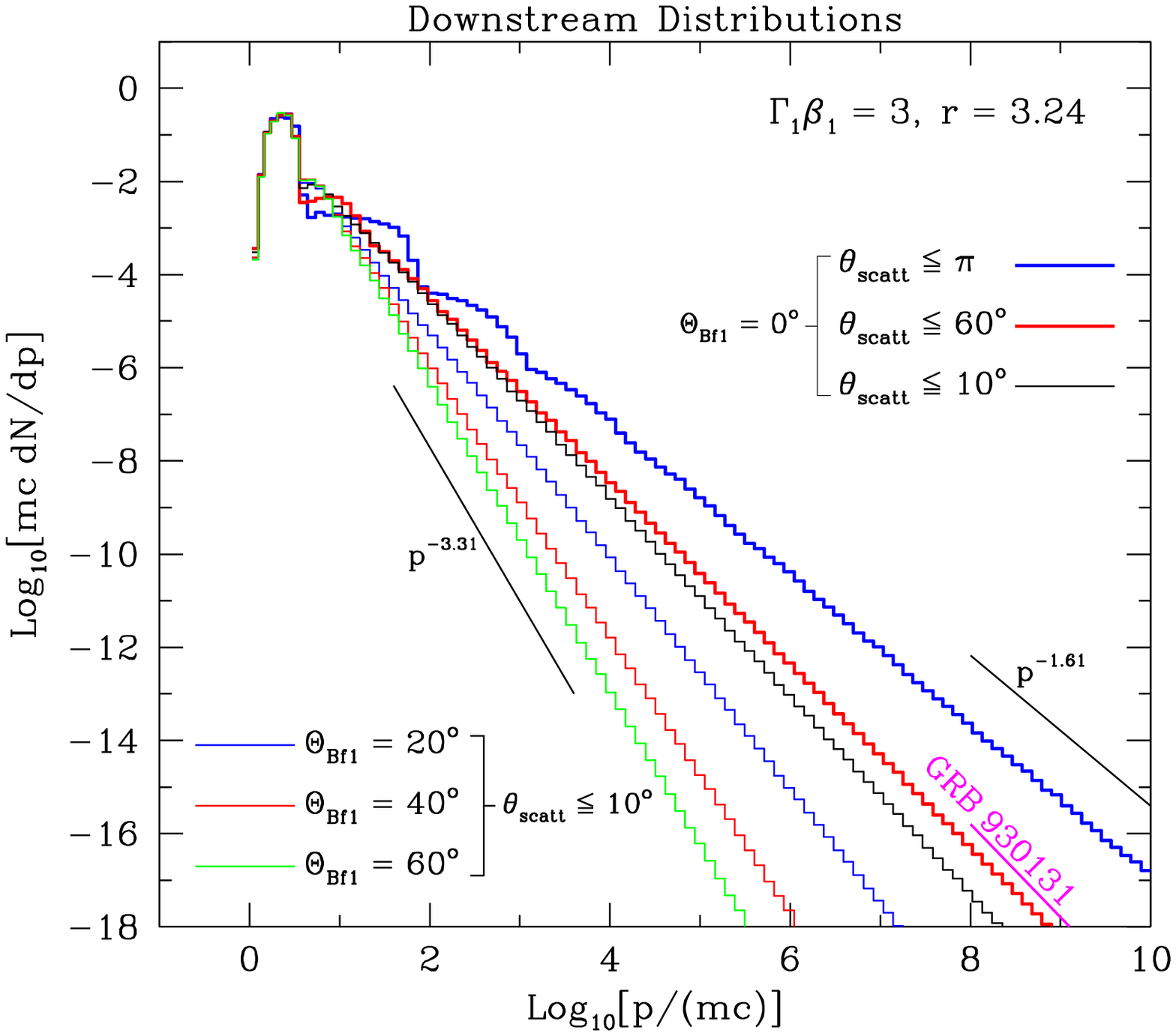}
   \hskip -0.1truecm
    \includegraphics[width=.51\textwidth]{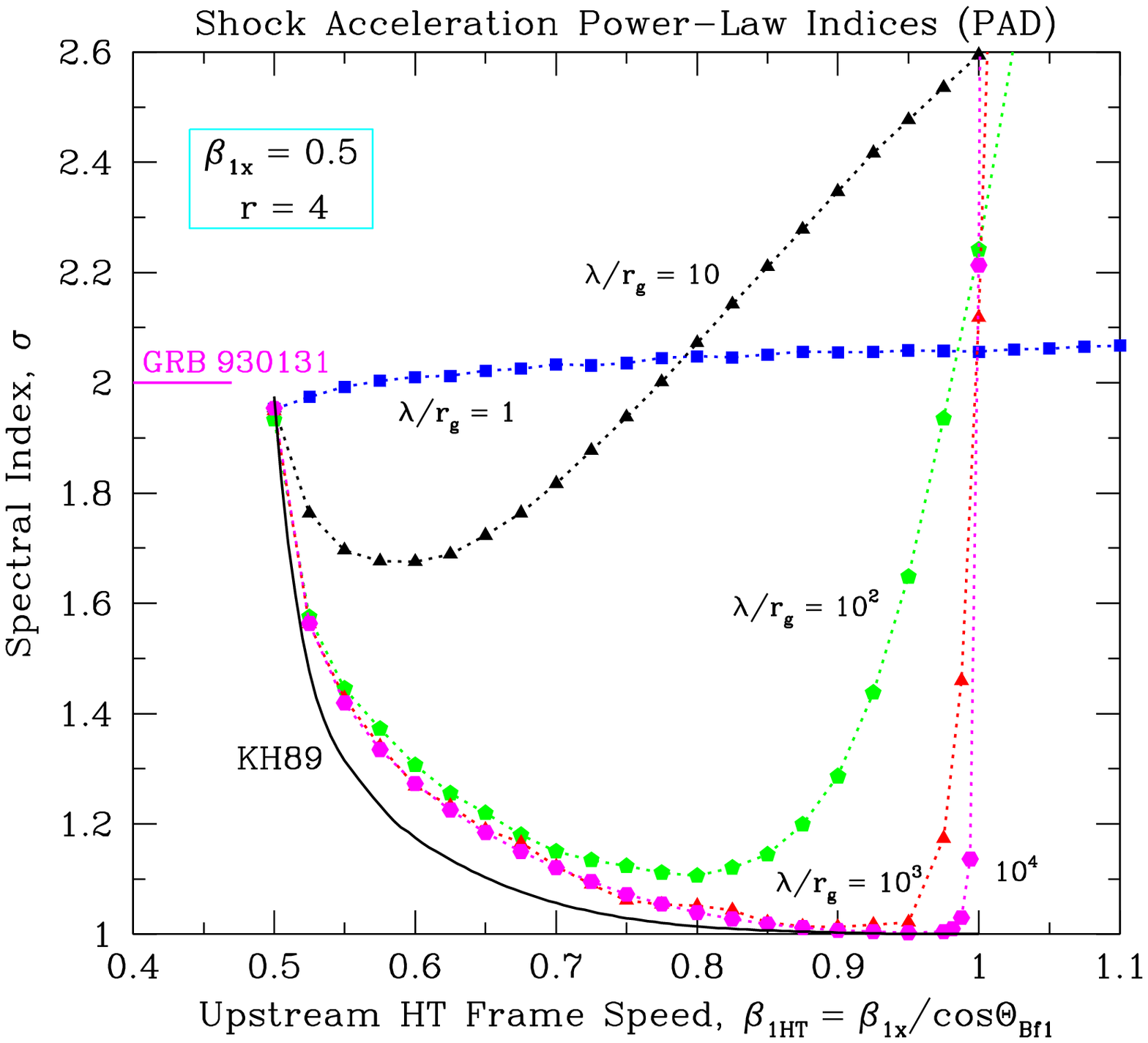}
  }
  \caption{{\it Left panel:}
Particle distribution functions  \teq{dN/dp} from mildly-relativistic
shocks (\teq{\Gamma_1\beta_1=3}, i.e. \teq{\beta_1=u_1/c=0.949}) of
upstream-to-downstream  velocity compression ratio
\teq{r=u_{1x}/u_{2x}\approx 3.24}. Simulation results can be divided
into two groups: parallel shock runs (\teq{\thetaBone=0^{\circ}}, upper
three histograms), and oblique, superluminal shock cases
(\teq{\thetaBone=20^{\circ}, 40^{\circ}, 60^{\circ}}, lower three
histograms).  Scattering off hydromagnetic turbulence was modeled by
randomly deflecting particle momenta by an angle within a cone, of
half-angle \teq{\thetascatt},  whose axis coincides  with the particle
momentum prior to scattering; the ratio of the diffusive mean free path
\teq{\lambda} to the gyroradius \teq{r_g} was fixed at \teq{\eta=\lambda
/r_g=5}. The heavyweight lines (two uppermost histograms) are for the
large angle scattering cases  (LAS: \teq{1/\Gamma_1\ll
\thetascatt\leq\pi}).  All other cases constitute pitch angle diffusion
(PAD) runs, when \teq{\thetascatt\lesssim 1/\Gamma_1} and the
distributions become independent of the choice of \teq{\thetascatt}. 
All distributions asymptotically approach power-laws \teq{dN/dp\propto
p^{-\sigma}} at high energies.  For the two cases bracketing the results
depicted, the power-laws are indicated by lightweight lines, with
indices of \teq{\sigma=1.61} (\teq{\thetaBone=0^{\circ}},
\teq{\thetascatt\leq\pi}) and \teq{\sigma =3.31}
(\teq{\thetaBone=60^{\circ}}, \teq{\thetascatt\leq 10^{\circ}}),
respectively.
\newline 
{\it Right panel:}
Power-law indices \teq{\sigma} for simulation runs in the limit of small
angle scattering (pitch angle diffusion), for an almost non-relativistic
shock of upstream flow speed \teq{\beta_{1x}\equiv u_{1x}/c =0.5}, and
an MHD velocity compression ratio \teq{r=4}. The indices are displayed
as functions of the effective de Hoffman-Teller frame upstream flow
speed \teq{\beta_{\hbox{\sevenrm 1HT}} = \beta_{1x}/\cos\thetaBone},
with select values of the fluid frame field obliquity \teq{\thetaBone}
marked at the top of the panel.  Obliquities for which
\teq{\beta_{\hbox{\sevenrm 1HT}} >1} constitute superluminal shocks. 
The displayed simulation index results were obtained for different
diffusive mean free paths \teq{\lambda} parallel to the mean field
direction, namely \teq{\lambda/r_g=1} (blue squares),
\teq{\lambda/r_g=10} (black triangles), \teq{\lambda/r_g=10^2} (green
pentagons), \teq{\lambda/r_g=10^3} (red triangles) and
\teq{\lambda/r_g=10^4} (magenta hexagons), as labelled.  The lightweight
black curve at the bottom labelled KH89 defines the semi-analytic result
from Kirk \& Heavens' \cite{KH89} solution to the diffusion-convection
equation, corresponding to \teq{\lambda/r_g\to\infty}.
\newline 
{\it Both panels:}
The short heavyweight magenta lines labelled GRB 930131 indicate the
approximate spectral index \teq{\sigma} that is appropriate for this
EGRET burst, if a cooled synchrotron emission scenario is operable (see
text for a discussion).
}
 \label{fig:laspad}
\end{figure}

Representative particle differential distributions \teq{dN/dp} that
result from the simulation of diffusive acceleration at
mildly-relativistic (internal GRB) shocks are depicted in the {\it left
panel} of Figure~\ref{fig:laspad} (see \cite{ED04,SBS07} for
\teq{\Gamma_1\gg 1} simulation results); these distributions are equally
applicable to electrons or ions, and so the mass scale is not specified.
A striking feature is that the slope and shape of the non-thermal
particle distribution depends on the nature of the scattering.  The
often cited asymptotic, ultrarelativistic index of \teq{\sigma =2.23}
for \teq{dN/dp\propto p^{-\sigma}} is realized only for parallel shocks
with \teq{\thetaBone =0^{\circ}} in the mathematical limit of small
(pitch) angle diffusion (PAD), where the particle momentum is
stochastically deflected on arbitrarily small angular (and therefore
temporal) scales. In practice, PAD results when the maximum scattering
angle \teq{\thetascatt} is inferior to the Lorentz cone angle
\teq{1/\Gamma_1} in the upstream region. In such cases, particles
diffuse in the region upstream of the shock only until their velocity's
angle to the shock normal exceeds around \teq{1/\Gamma_1}, after which
they are rapidly swept downstream of the shock. The Figure indicates
clearly that when the field obliquity increases, so also does the index
\teq{\sigma}, with values greater than \teq{\sigma\sim 3} arising for
\teq{\thetaBone\gtrsim 50^{\circ}} for this mildly-relativistic
scenario.  This is a consequence of more prolific convection downstream
away from the shock.

Figure~\ref{fig:laspad} also shows results for large angle scattering
scenarios (LAS, with \teq{4/\Gamma_1\lesssim \thetascatt\lesssim\pi}),
where the distribution is highly structured and much flatter on average
than \teq{p^{-2}}.  The structure becomes more pronounced for large
\teq{\Gamma_1} \cite{Baring04,ED04,SBS07}, and is kinematic in origin,
where large angle deflections lead to fractional energy gains between
unity and \teq{\Gamma_1^2} in successive shock crossings.  Each
structured bump or spectral segment corresponds to an increment of two
in the number of shock transits.  For \teq{p\gg mc}, they asymptotically
relax to a power-law, in this case with index \teq{\sigma\approx 1.62}. 
Intermediate cases are also depicted in Figure~\ref{fig:laspad},  with
\teq{\thetascatt\sim  4/\Gamma_1}.  The spectrum is  smooth, like for
the PAD case, but the index is lower than 2.23.  From the plasma physics
perspective, magnetic turbulence could easily be sufficient  to effect
scatterings on this intermediate angular scale, a contention that 
becomes even more germane for ultrarelativistic shocks \cite{SBS07}.
Astrophysically, such flat distributions may prove desirable in models
of GRB or blazar jet emission, particularly if efficient radiative
cooling is invoked.  Note that acceleration distributions are
qualitatively similar for ultra-relativistic shocks, i.e. those perhaps
more appropriate for GRB afterglow studies, with a trend
\cite{EJR90,Kirk00,Baring04,ED04,SBS07} of declining \teq{\sigma} for
higher \teq{\Gamma_1}, the consequence of an increased kinematic energy
boosting in collisions with turbulence.

It is clear that there is a considerable range of indices \teq{\sigma}
possible for non-thermal particles accelerated in mildly relativistic
shocks, a non-universality that can attractively mesh with GRB
observations.. The two key parameters that dictate the array of possible
spectral indices \teq{\sigma} are the shock field obliquity
\teq{\thetaBone}, and the ratio \teq{\eta =\lambda /r_g}of the diffusive
mean free path to the particle's gyroradius \teq{r_g}.  A parameter
survey for diffusive acceleration at a typical mildly-relativistic shock
is exhibited in the {\it right panel} of Fig.~\ref{fig:laspad}, where
only the pitch angle diffusion limit was employed. The power-law index
\teq{\sigma} is plotted as a function of the de Hoffman-Teller (HT
\cite{dHT50}) frame dimensionless speed \teq{\betaHTone
=\beta_{1x}/\cos\thetaBone}, where the subscript \teq{x} denotes
components along the shock normal. This is the upstream flow speed in
the shock rest frame where the flow is everywhere parallel to the
magnetic field; it corresponds to a physical speed when \teq{\betaHTone
< 1} (i.e. the upstream field obliquity satisfies \teq{\cos\thetaBone <
\beta_{1x}}, and the shock is said to be {\it subluminal}.  When
mathematically \teq{\betaHTone > 1}, the shock is termed {\it
superluminal}, and the de Hoffman-Teller frame technically does not
exist --- no Lorentz boost can render the flow parallel to {\bf B}. 
Clearly, the x-axis maps variations in \teq{\thetaBone}, with more
highly oblique shocks corresponding to the right of the plot.

A feature of this plot is that the dependence of \teq{\sigma} on field
obliquity is non-monotonic.  When \teq{\lambda /r_g\gg 1}, the value of
\teq{\sigma} at first declines as \teq{\thetaBone} increases above zero.
This leads to very flat spectra.  As \teq{\betaHTone} approaches and
eventually exceeds unity, this trend reverses, and \teq{\sigma} then
rapidly increases with increasing shock obliquity.  This is the
character of near-luminal and superluminal shocks evinced in the left
panel of Fig.~\ref{fig:laspad}, precipitated by inexorable convection of
particles away downstream of the shock, steepening the distribution
dramatically. The only way to ameliorate this rapid decline in the
acceleration efficiency is to reduce \teq{\lambda /r_g} to values below
around \teq{10}.  Physically, this is tantamount to increasing the
hydromagnetic turbulence to high levels that force the particle
diffusion to approach isotropy.  Then, transport of charges across the
mean field becomes significant on gyrational timescales, and they can be
retained near the shock for sufficient times to accelerate and generate
suitably flat distribution functions.  Such low values of \teq{\lambda
/r_g} render the field direction immaterial, and the shock behaves much
like a parallel, subluminal shock in terms of its diffusive character.
Then, \teq{\sigma} is only weakly dependent on \teq{\thetaBone}, the
second key property illustrated in the right panel.

The origin of the extremely flat distributions with \teq{\sigma\sim 1}
is in the coherent effect of {\it shock drift acceleration} at the shock
discontinuity.  This phenomenon has been widely understood in
non-relativistic shocks, and is due to the energy gain of charges when
they repeatedly encounter {\bf u}\teq{\times}{\bf B} electric fields (in
frames other than the HT frame) in gyrations straddling the shock
discontinuity.  The energy gain rate then scales as the gyrofrequency
\teq{c/r_g}, i.e. is proportional to the particle's energy/momentum, so
that the equilibrium distribution realized in pure shock drift
acceleration is \teq{dN/dp\propto p^{-1}} \cite{BS09}.  Reducing
\teq{\lambda /r_g} and thereby introducing extremely modest amounts of
cross-field diffusion disrupts this coherence, removes particles from
the shock layer, and steepens the spectrum. It is noted in passing that
the choice of the somewhat large (for \teq{\beta_{1x}=0.5}) compression
ratio \teq{r=4} was to afford direct comparison with the semi-analytic
convection-diffusion equation results of Kirk \& Heavens \cite{KH89},
which physically correspond to \teq{\lambda /r_g\gg 1} cases with
minimal cross field diffusion. The differences between our simulation
indices in this limit and those of \cite{KH89} arise because the
adiabatic approximation employed in \cite{KH89} to describe shock
crossings does not take into account the gyrational trajectories of
charges in the shock layer.  These subtleties are discussed further in
\cite{BS09}.

\section{Constraints on Acceleration Theory from GRB Observations}
 \label{sec:obs_vs_theory}

Gleaning meaningful insights into the connection between acceleration
theory and prompt burst signals hinges upon two aspects: the high energy
spectral index \teq{\alpha_h} (\teq{=-\beta} for Band model
\cite{Band93} index \teq{\beta}), when the differential spectrum is
\teq{dn/d\erg\propto \erg^{-\alpha_h}}, and generating burst spectral
shape at and below the \teq{\nu}-\teq{F_{\nu}} peak.  These will be
considered in turn.

\subsection{The \teq{\alpha_h} High-Energy Spectral Index}

The high-energy photon spectral indices \teq{\alpha_h} for the CGRO
EGRET and Comptel bursts are tabulated in \cite{Baring06}. One of these
EGRET detections, GRB 930131, had an index \teq{\alpha_h \approx 2},
which corresponds to the indication in both panels of
Fig.~\ref{fig:laspad} if a synchrotron or inverse Compton cooling model
is adopted.  This is on the flat end of the EGRET \teq{\alpha_h} index
distribution \cite{Dingus95}, for which a handful of sources evinced
photon differential spectrum indices scattered in the range
\teq{2\lesssim\alpha_h\lesssim 3.7} (though note that EGRET spark
chamber detections were concentrated in the index range \teq{\alpha_h
\lesssim 2.8} \cite{Baring06}).  The CGRO BATSE  \teq{\alpha_h} index
distribution \cite{Preece00} is similarly broad, but with
\teq{1.5\lesssim \alpha_h\lesssim 3.5} and far greater statistics.   The
recent {\it Fermi} detection of GRB 08016C in both the GBM and LAT
instruments offered a high energy index of \teq{\alpha_h\sim 2.2} in its
most luminous epoch, and a steeper spectrum at other times
\cite{Abdo09}.  Accordingly, observationally, shock acceleration models
must accommodate a radiation spectral index in the range
\teq{2\lesssim\alpha_h\lesssim 4} in order to be viable.

The phase space for achievability of this depends on the radiation
mechanism, and also upon whether or not cooling is invoked.  The most
popular model scenario for the production of the prompt GRB signal is
quasi-isotropic electron synchrotron emission \cite{Piran99,Meszaros02}.
In this case, as with inverse Compton scattering, the photon
differential spectral index is given by \teq{\alpha_h = (\sigma + 1)/2}
above the \teq{\nu}-\teq{F_{\nu}}  peak, if cooling is not efficient,
and \teq{\alpha_h = (\sigma + 2)/2} if the particles are continuously
cooled in a large volume before being re-injected into an acceleration
process, for which the particle distribution index is steepened by
unity.   From the Figure, it is evident that this is readily achievable
for GRB 930131 for quasi-parallel shocks with perhaps some component of
modest angle scattering, or for moderately oblique shocks as long as
\teq{\lambda /r_g\lesssim 3-10}.  The EGRET GRB index range
\teq{2\lesssim\alpha_h\lesssim 3.7} then corresponds to diffusive
acceleration with \teq{2\lesssim\sigma\lesssim 5.4} in the presence of
cooling, and \teq{3\lesssim\sigma\lesssim 6.4} otherwise.  In either
case, subluminal shocks, either quasi-parallel (\teq{\thetaBone\lesssim
20^{\circ}}) or moderately oblique
(\teq{20^{\circ}\lesssim\thetaBone\lesssim 60^{\circ}}) are required to
explain flatter GRB spectra as in GRB 910503, GRB 930131, GRB 950425 
or GRB 080916C. For bursts with somewhat larger \teq{\alpha_h} indices,
superluminal shocks can be viable, but only if the scattering is strong,
i.e. \teq{\lambda /r_g\lesssim 3-10}. Otherwise, convection away from
the shock dominates the acceleration, and the electron distribution
becomes too steep.  For the less popular uncooled hadronic models, the
photon and particle indices trace each other, i.e. \teq{\alpha_h =
\sigma}, and a similar conclusion that the environment is restricted to
subluminal or modestly superluminal oblique shocks is derived.  Note
that while Lorentz transformations of circumburst fields of arbitrary
orientation render ultra-relativistic external GRB shocks
quasi-perpendicular (\teq{\thetaBone\approx 90^{\circ}}), moderate field
obliquities are often attained in mildly-relativistic internal shocks.

Discriminating between the \teq{\sigma} constraints in the presence or
absence of synchrotron cooling mandates a brief discussion of the
balance between acceleration and cooling.  Diffusive shock acceleration
is usually dominated by gyroresonant interactions between charges and
hydromagnetic turbulence \cite{Drury83}. If this is the case, then for a
charge of Lorentz factor \teq{\gamma}, the acceleration rate essentially
scales as \teq{\gamma} times the particle's gyrofrequency, implying
\teq{d\gamma/dt\propto \gamma^0}. In contrast, synchrotron and inverse
Compton cooling rates yield \teq{d\gamma/dt\propto -\gamma^2}. 
Accordingly, simultaneous and co-spatial acceleration and synchrotron
cooling yield a power-law distribution with the ``raw'' acceleration
index \teq{\sigma} up to a maximum energy where cooling begins to
dominate, and generates an exponential turnover.  This is the widely
considered scenario for X-ray/TeV supernova remnant studies.  For
external GRB shocks spawning afterglow signals, there is plenty of time
for accelerated charges to diffuse away from the shock and subsequently
cool, so that the cumulative effect is emission from a cooling-steepened
injected power-law.  The situation is very different for internal
shocks.  The acceleration must persist up to the maximum energy seen in
the observations, for example 1 GeV in GRB 930131 and 13 GeV in GRB
080916C, and on a timescale \teq{\Delta t} commensurate with the
variability at these energies, i.e., \teq{\Delta t\lesssim 1}sec or much
less.  Cooling breaks, by an index of \teq{\Delta\alpha_h =1/2}, are not
observed above 1 MeV in burst emission, so if the spectra spanning 1 MeV
-- 10 GeV correspond to a cooling-dominated contribution, the
acceleration must operate in short, impulsive periods, followed by long
cooling epochs that permit the electrons to decline in momentum by a
factor of \teq{10^2} or so. Accordingly, the cooling epochs should
possess durations of around 4 orders of magnitude longer than the
impulsive acceleration epochs.  It is not clear that the superposition
of a host of these presumably pseudo-FRED-like profiles can mimic
typical burst time histories while maintaining the time-averaged
spectral behavior.

\subsection{Constraints from the \teq{\nu}-\teq{F_{\nu}} Peak}

An additional requirement for the success of an acceleration model is
that when convolved with a radiation process such as synchrotron
radiation, it can generate the shape of the GRB spectrum around and
below the \teq{\nu}-\teq{F_{\nu}} peak.  This was the focus of the
investigation by Baring \& Braby \cite{BB04}.  Early work on broad-band
spectral fitting of bursts was provided by Tavani \cite{Tavani96}, who
obtained impressive fits to time-integrated spectra of several bright
CGRO BATSE bursts using a phenomenological electron distribution and the
synchrotron emission mechanism.  While this succeeded in this limited
sample, providing a driver for the popularity of the synchrotron GRB
emission model, there are difficulties with fitting low energy (i.e.
\teq{\lesssim 100}keV) spectra in about 1/3 of bursts \cite{Preece98}. 
Employing perspectives based on acceleration theory, this investigative
effort was extended and refined by \cite{BB04}, who pursued a program of
spectral fitting of GRB emission using a linear combination of thermal
and non-thermal electron populations. These fits demanded that the
preponderance of electrons that are responsible for the prompt emission
reside in an intrinsically non-thermal population. Such a constraint,
dictated by the narrowness of the \teq{\nu}-\teq{F_{\nu}} peak, is
commensurate with electron distributions frequently presumed in GRB
spectral models, i.e. truncated power-laws.  Yet, this requirement
strongly contrasts particle distributions obtained from acceleration
simulations, exemplified by those depicted in Fig.~\ref{fig:laspad}, and
further illustrated in numerous papers
\cite{EJR90,EBJ95,ED04,SB06,SBS07}. Moreover, particle-in-cell (PIC)
plasma simulations \cite{Hoshino92,Nishikawa05,Medvedev05} generally
exhibit largely Maxwellian distributions, and in the isolated recent
suggestion \cite{Spitkovsky08} of a non-thermal tail generated by
diffusive transport, the thermal population strongly dominates the
high-energy tail.

The consequence is obviously a potential conflict for acceleration
models, since the non-thermal electrons are almost always drawn directly
from a thermal gas. This does not necessarily mean that acceleration at
relativistic shocks is not precipitating the prompt emission.  It is
possible that somehow, relativistic shocks can suppress thermalization
of electrons, though such a conjecture presently has no simulational
evidence to support it. Or, radiative efficiencies might become
significant only at highly superthermal energies: a convenient
resolution to this dilemma is that strong radiative self-absorption
could be acting, in which case \teq{\nu}-\teq{F_{\nu}} peaks (and
therefore BATSE and {\it Fermi} GBM spectral probes) are not actually
sampling thermal electrons.  It is also possible that other mechanisms
such as pitch-angle synchrotron, or jitter radiation may prove more
desirable.  A cautionary note is that this complication for the
interpretation is predicated upon spectra integrated over the burst
duration, which can differ significantly from those obtained in temporal
sub-intervals (e.g. see \cite{Abdo09}).  Notwithstanding, cumulative
spectra provide a global indication of a narrowness of the MeV spectral
break, and this must be a characteristic of at least the most luminous
portions of the burst time history.  We note also that \cite{BB04}
concluded that the synchrotron self-Compton process was unlikely to be
able to generate the MeV band emission in prominent EGRET bursts because
of its intrinsic spectral breadth, even with extremely narrow electron
distribution functions.  Only a self-absorption in the synchrotron seed
spectrum can aid its viability.

\section{Conclusion}

This paper explores the connection between diffusive shock acceleration
theory and prompt emission in bursts.  Simulation results presented
clearly highlight the non-universality of the index of energetic,
non-thermal electrons and ions, spawned by the variety of shock
obliquities and the character of hydromagnetic turbulence in their
environs.  This non-universality poses no problem for modeling GRB
high-energy power-law indices, though observations generally constrain
the parameter space to subluminal or highly-turbulent superluminal
shocks near the Bohm diffusion limit.  Also, demanding a connection
between the thermal electron population and the GRB
\teq{\nu}-\teq{F_{\nu}} peak gives rise to an inconsistency in that
acceleration simulations predict a dominance of the thermal population,
whereas the radiation models demand that the contributing particles are
intrinsically non-thermal.  Escape clauses include invoking radiative
self absorption, or perhaps relinquishing preferred emission mechanisms,
i.e., synchrotron or inverse Compton radiation. The prospect of a number
of platinum standard, broad-band GRB detections by {\it Fermi} that
permit time-dependent spectroscopy should hone our understanding of the
connection between shock acceleration and prompt emission.
\newline
{\bf Acknowledgments:} this research was supported in part by NASA grant
NNG05GD42G and NSF grant 0758158.

\bibliographystyle{aipproc}

\end{document}